\documentclass[useAMS]{gGAF2e}
\usepackage{graphicx, subfigure}
\usepackage[draft]{changes}

\renewcommand*{\Re}{{\mathrm {Re}}}
\renewcommand*{\hat}{\widehat}

\begin{document}
\markboth{X.~Wei}{Kinematic dynamo induced by helical waves}
\title{Kinematic dynamo induced by helical waves}
\author{Xing Wei\thanks{Email: xingwei@astro.princeton.edu}\\Princeton University Observatory, Princeton, NJ 08544, USA
\\\vspace{6pt}\received{Received 20 February 2014; in final form 30 May 2014; first published online ????}}
\date{\today}
\maketitle

\begin{abstract}
We investigate numerically the kinematic dynamo induced by the superposition of two helical waves in a periodic box as a simplified model to understand the dynamo action in astronomical bodies. The effects of magnetic Reynolds number, wavenumber and wave frequency on the dynamo action are studied. It is found that this helical-wave dynamo is a slow dynamo. There exists an optimal wavenumber for the dynamo growth rate. A lower wave frequency facilitates the dynamo action and the oscillations of magnetic energy emerge at some particular wave frequencies.\\

\end{abstract}

\section{Introduction}
Magnetic fields in astronomical bodies, e.g., planets, stars, galaxies and disks, are believed to be generated by the dynamo action, the motion of electrically conducting fluid shearing and twisting field lines to create new lines through the electromagnetic induction effect to offset magnetic diffusion. In the natural dynamo action, rotation plays an important role because the Coriolis force leads to a helical spatial structure of fluid flow and this helical flow facilitates the dynamo action through the so-called $\alpha$-effect, i.e., the fluid helical motion twisting field lines. In convection driven dynamo, either the large-scale columnar rolls in the Earth's core or the small-scale turbulent eddies in the solar interior have the helical structure \citep{busse,zhang_schubert,glatzmaier_roberts,jones}. In mechanical force driven dynamo, e.g., dynamo driven by Couette flow or libration or precession or tide, the inertial waves arising from the destabilisation of internal shear layers also have the helical structure \citep{tilgner1,cardin,wei}. In rotating turbulence with the lack of reflexional symmetry, the first order smoothing result shows that the turbulent electromotive force (e.m.f.) is proportional to the local mean field and the coefficient tensor is correlated to fluid helicity, i.e., the scalar product of velocity and vorticity. This result was independently discovered by \citet{steenbeck} and \citet{moffatt_kinematic}. Moreover, in \cite{moffatt_kinematic} the correlation between the coefficient tensor and the turbulent helicity spectrum was physically interpreted with helical waves. In \cite{moffatt_dynamic} the dynamics of fluid motion with a uniform rotation was considered and the dynamo induced by superposition of random inertial waves was analytically studied in both linear and weakly nonlinear regimes.

In this short paper, we do not involve the fluid dynamics but focus on the kinematic dynamo induced by helical waves \citep[the kinematic dynamo induced by drifting wave was studied by][]{tilgner2}. The fluid motion of one helical wave is planar (see the next section) and the Zel'dovich anti-dynamo theorem states that a planar flow cannot generate the dynamo action. Therefore we use the superposition of two helical waves to induce the dynamo action, which is a very simple model of the $\alpha^2$-dynamo. An interesting kinematic dynamo is the fast dynamo, in which magnetic field grows on the advection time scale and the growth rate is independent of the magnetic diffusion time scale. That is, the growth rate of a fast dynamo is finite when the magnetic Reynolds number approaches infinity \citep{soward,childress}. For example, the Roberts flow, a steady two-dimensional flow in a periodic box, generates a slow dynamo \citep{roberts} whereas the Galloway-Proctor flow, an oscillatory two-dimensional flow in a periodic box, generates a fast dynamo \citep{galloway_proctor} (the sense of `two-dimensional' is in respect of two coordinates but not two components), because ``the Roberts flow has no exponential separation of nearby points but Galloway-Proctor flow has large stretching in large regions of the domain'' \citep{proctor}. Another example of fast dynamo in spherical geometry was given in \cite{hollerbach}. Whether the helical-wave dynamo is a fast dynamo is one of the purposes of this study. The other two purposes are to study the effects of wavenumber and wave frequency on the growth rate of dynamo. In section 2 the problem is formulated, in section 3 the results are shown, and in section 4 a short discussion is given.

\section{Formulation}
We use the Cartesian coordinate system $(x_1,x_2,x_3)$. Suppose that we have a plane wave $\bm u$ traveling in the $x_1$-direction with wave vector $\bm k=k{\hat {\bm x}_1}$, frequency $\varpi$ and complex amplitude $\hat{\bm u}$, i.e., $\bm u=\Re\{\hat{\bm u}\exp[{\mathrm i}(kx_1-\varpi t)]\}$.
For a helical wave satisfying $\bm\nabla\times\bm u=k\bm u$, the three components of $\hat{\bm u}$ are given to be
\begin{equation}\label{wave}
\hat{\bm u}=(0,\,\hat u_0,\,{\mathrm i}\hat u_0),
\end{equation}
which indicates that the fluid motion induced by a helical wave has no component in the direction of phase propagation but is circular in the plane perpendicular to phase velocity, namely the helical wave is a transverse wave. We denote the wave amplitude $|\bm u|=|\Re\{\hat{\bm u}\}|=|\hat u_0|$ by $u_0$. The helicity of a helical wave is then $h=\bm{ u\,\cdot\,}(\bm\nabla\times\bm u)=ku^2_0$.

We now consider the dynamo action in a periodic box. The dimensionless magnetic induction equation reads
\begin{equation}\label{induction}
\frac{\upartial\bm B}{\upartial t}=\bm\nabla\times(\bm u\times\bm B)+Rm^{-1}\nabla^2\bm B,
\end{equation}
where the characteristic length is taken to be the box size $l$, the characteristic velocity to be the wave amplitude $u_0$, and the characteristic time to be the advection time $l/u_0$. The magnetic Reynolds number
\begin{equation}
Rm=\bigl.{u_0l}\bigr/{\eta},
\end{equation}
where $\eta$ is the magnetic diffusivity, measures the induction effect against the magnetic diffusion. It should be noted that in \cite{moffatt_kinematic, moffatt_dynamic} the characteristic length was taken to be the wavelength $\sim1/k$ but not the box size such that dynamo occurs at a fairly low $Rm_{\rm wave}$, corresponding to small-scale waves. If a uniform field $\bm B_0$ is imposed in the $x_1$-direction then the electromotive force induced by this helical wave can be readily derived from the first-order perturbation of magnetic induction equation, i.e.,
\begin{equation}\label{emf}
\bm u\times\bm b\,=\,-\,\frac{\eta u^2_0B_0k^3}{\varpi^2+\eta^2k^4}\hat{\bm x}_1\,=\,-\,\frac{\eta B_0k^2}{\varpi^2+\eta^2k^4}h\hat{\bm  x}_1,
\end{equation}
which indicates that the magnetic diffusivity is essential to the electromotive force. The detail of this derivation can be found in \cite{moffatt_book}. This expression will be used to interpret our numerical results in the next section. It also implies that the imposed field and the induced current have the opposite directions for a right-handed helical wave while they are in the same direction for a left-handed helical wave.

The helical-wave dynamo is mathematically a linear eigenvalue problem and in principle it can be solved with the Floquet analysis \citep{soward}. However, in the Floquet analysis the coefficient matrix is huge and the numerical calculations are demanding. Therefore we use the time-stepping method to numerically solve the magnetic induction equation (\ref{induction}) with a pseudo-spectral code in which the fast Fourier transform is done back and forth for the calculation of the induction term and the diffusion term is treated implicitly for numerical stability. A random initial field is given such that all the possible eigenmodes are involved and then the equation will select the fastest growing mode. Because we do not know which mode will be the fastest growing one in the dynamo action we calculate the total magnetic energy to judge the onset of dynamo instability. Suppose that the magnetic field is expressed as $\bm B(\bm x,t)=\Re\{\hat{\bm B}(\bm x)\exp(\sigma t)\}$ where $\hat{\bm B}$ is the complex amplitude and $\sigma$ the complex growth rate. The total magnetic energy can then be calculated as
\begin{align}
\int_V&\tfrac12 {B^2}dV\nonumber\\[-0.6em]
&=\tfrac{1}{4}\exp(2\sigma_r t)\int_V\left[\bigl(\hat{B}_r^2+\hat{B}_i^2\bigr)+\bigl(\hat{B}_r^2-\hat{B}_i^2\bigr)\cos(2\sigma_it)-2\hat{\bm B}_r{\,\bm \cdot\,}\hat{\bm B}_i\sin(2\sigma_it)\right]{\mathrm d}V,\qquad\quad\label{Em}
\end{align}
where the subscripts $r$ and $i$ denote respectively the real and imaginary parts. Equation (\ref{Em}) shows that the magnetic energy depends not only exponentially but also harmonically on time.

In most of numerical calculations in this paper we use the two helical waves, i.e.,
\begin{equation}\label{u}
\bm u=\bm u_1+\bm u_2=\Re\bigl\{\hat{\bm u}_1e^{{\mathrm i}(k_1x_1-\varpi_1 t)}\bigr\}+\Re\bigl\{\hat{\bm u}_2e^{{\mathrm i}(k_2x_2-\varpi_2 t)}\bigr\},
\end{equation}
where
\begin{equation}\label{u1u2}
\hat{\bm u}_1=\bigl(0,\,\hat u_0,\,{\mathrm i}\hat u_0\bigr) \hspace{15mm}\text{and}\hspace{15mm} \hat{\bm u}_2=\bigl({\mathrm i}\hat u_0,\,0,\,\hat u_0\bigr).
\end{equation}
Because our computational domain is chosen to be $[0,2\pi]^3$, the wavenumbers $k_1$ and $k_2$ are integers to keep the periodicity. The flow $\bm u$ is two-dimensional in the sense that it depends on two coordinates $x_1$ and $x_2$. Alternatively, we can also choose the flow to depend either on $x_2$ and $x_3$ or on $x_3$ and $x_1$, but our numerical calculations proved that this cannot influence the growth rate of dynamo. It has also been proved that the phase of $\hat u_0$ cannot influence the growth rate either, but the amplitude $u_0$, the wavenumbers $k_1$ and $k_2$, and the wave frequencies $\varpi_1$ and $\varpi_2$ can. In most of calculations we choose the identical wavenumbers and wave frequencies, i.e., $k_1=k_2$ and $\varpi_1=\varpi_2$.

\section{Results}

\begin{figure}
\centering
\includegraphics[scale=0.6]{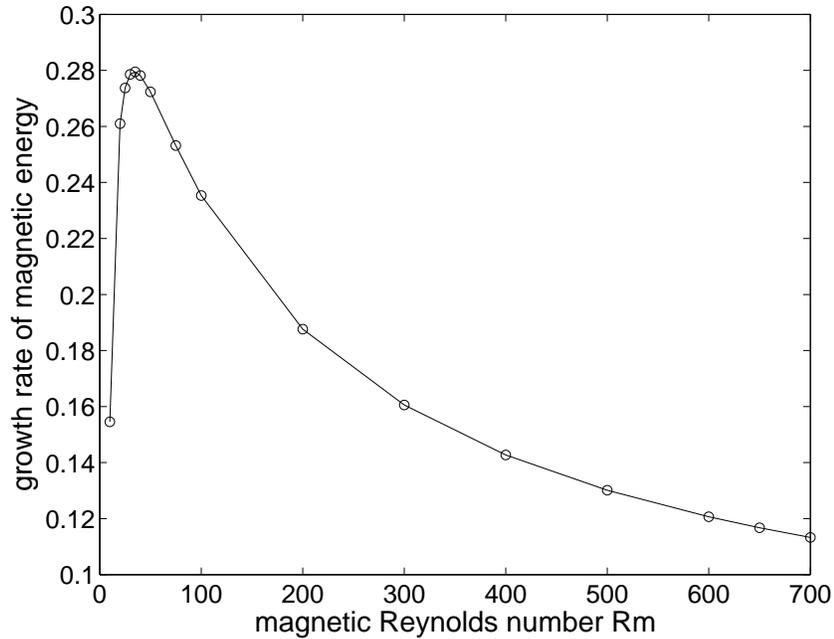}
\caption{Growth rate of magnetic energy against $Rm$ at $k_1=k_1=1$ and $\varpi_1=\varpi_2=0.01\times2\pi$. Circle symbol denotes $\sigma_i=0$.}
\label{fig1}
\end{figure}

This dynamo induced by two helical waves is an $\alpha^2$-dynamo, in which magnetic field is twisted by two helical waves (one needs to note that magnetic field is always three-dimensional). The first test is about $Rm$. Figure \ref{fig1} shows the growth rate of magnetic energy against $Rm$. It clearly shows that the growth rate increases and then decreases with the increasing $Rm$. $Rm=35$ corresponds to the maximum growth rate. When $Rm$ is large enough the growth rate decreases to zero, namely the helical-wave dynamo is a slow dynamo. As pointed out by \cite{moffatt_proctor} that fast dynamos with smooth (differentiable) magnetic fields do not exist, the helical-wave dynamo is a slow dynamo because the flow stretching caused by the helical-wave motion is not efficient enough such that magnetic field cannot grow on the fast advection time scale at large $Rm$. On this curve, the two wavenumbers are $1$ and the two frequencies are $0.01\times2\pi$. The magnetic energy has only the exponential growth but not the oscillation, i.e., $\sigma_i=0$ in (\ref{Em}). The curves at other wavenumbers and frequencies are similar, only that $Rm$ corresponding to the maximum growth rate is different. The magnetic energy is always on a large scale in both the $x_1$ and $x_2$-directions but has a smaller scale in the $x_3$-direction when $Rm$ becomes higher.

\begin{figure}
\centering
\subfigure[$Rm=20$]{
\includegraphics[scale=0.6]{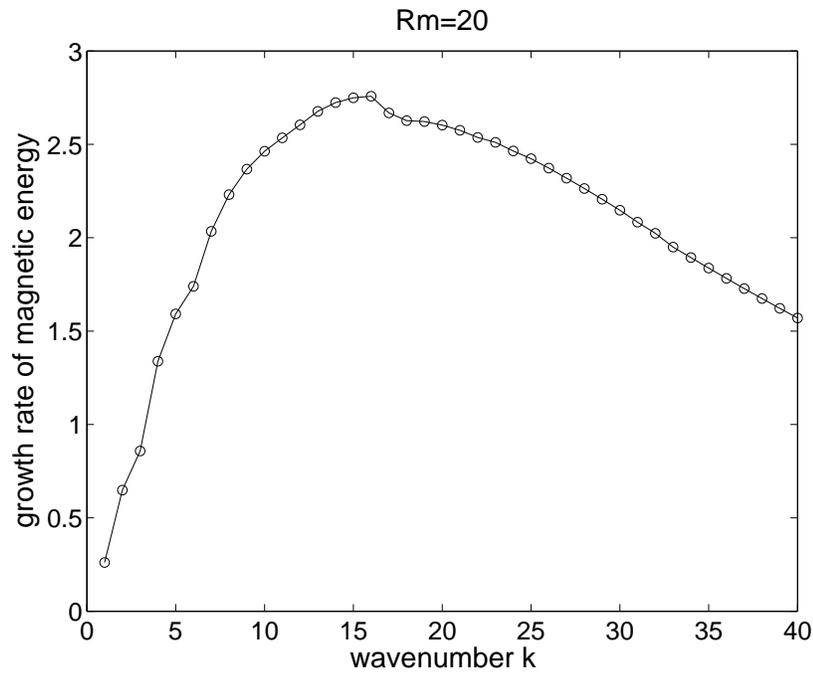}
\label{fig2a}}
\subfigure[$Rm=50$]{
\includegraphics[scale=0.6]{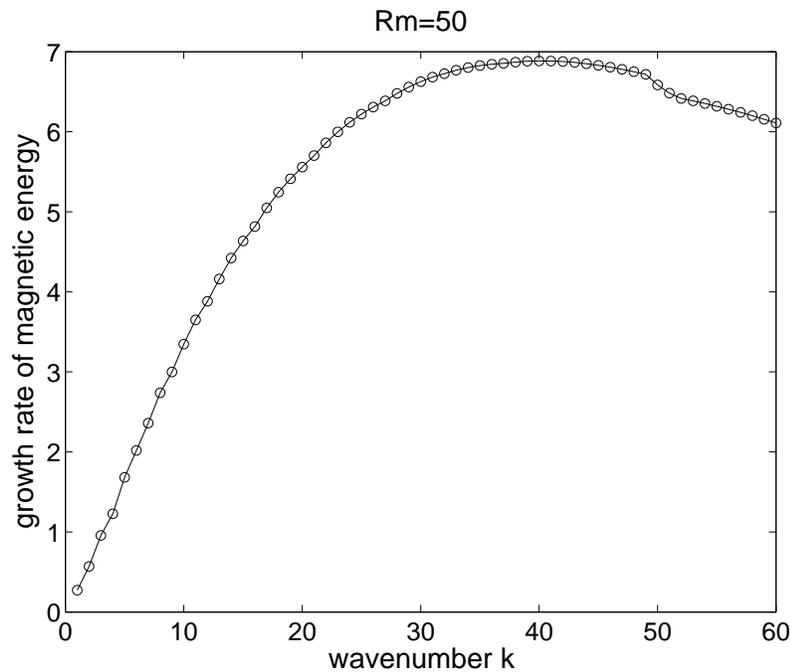}
\label{fig2b}}
\caption{Growth rate of magnetic energy against wavenumber $k=k_1=k_2$ at $\varpi_1=\varpi_2=0.01\times2\pi$. (a) $Rm=20$ and (b) $Rm=50$. Circle symbol denotes $\sigma_i=0$.}
\label{fig2}
\end{figure}

Then we test the effect of wavenumber on the growth rate. Figure \ref{fig2} shows the growth rate of magnetic energy against wavenumber. Again, the magnetic energy has only exponential growth, i.e., $\sigma_i=0$. Both the subfigures at two different $Rm$'s indicate that there exists an optimal wavenumber corresponding to the maximum growth rate. This can be qualitatively interpreted with equation (\ref{emf}) which indicates that there exists an optimal wavenumber for the $\alpha$-effect. One may notice that the optimal wavenumber deduced from (\ref{emf}) depends on frequency and diffusivity but not on $u_0$ or $Rm$ in the dimensionless form, and this is contradictory to the numerical result as shown in figure \ref{fig2}, namely the two different $Rm$'s correspond to the two different optimal wavenumbers. This is because equation (\ref{emf}) is for the $\alpha$-effect induced by the interaction of one helical wave and an externally imposed field, but in the helical-wave dynamo the flow is the superposition of two helical waves, which leads to the cross terms for the $\alpha$-effect, and the magnetic field is not externally imposed but self-excited (see the last paragraph in this section for details).

\begin{figure}
\centering
\subfigure[$Rm=20$]{
\includegraphics[scale=0.6]{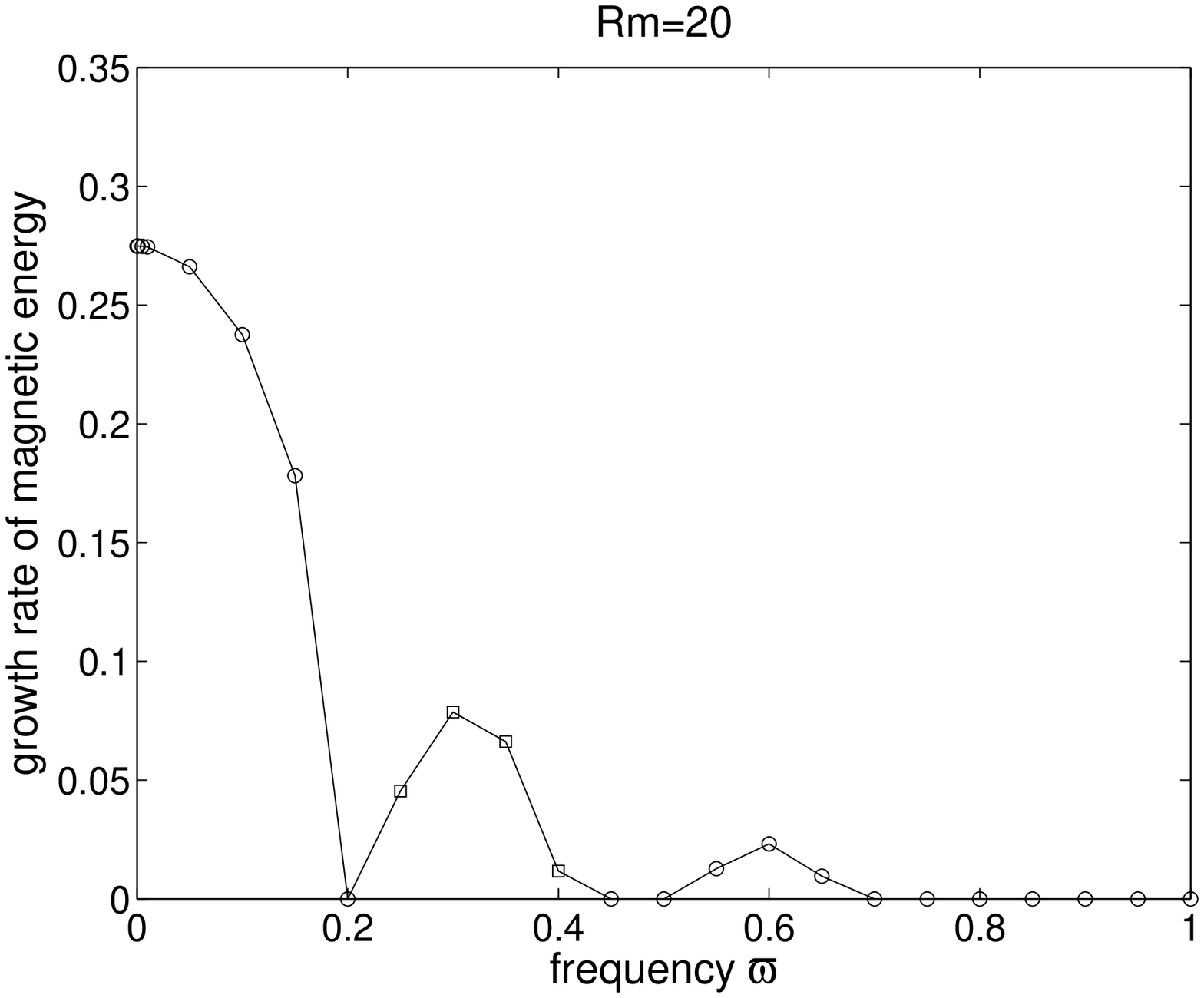}
\label{fig3a}}
\subfigure[$Rm=50$]{
\includegraphics[scale=0.6]{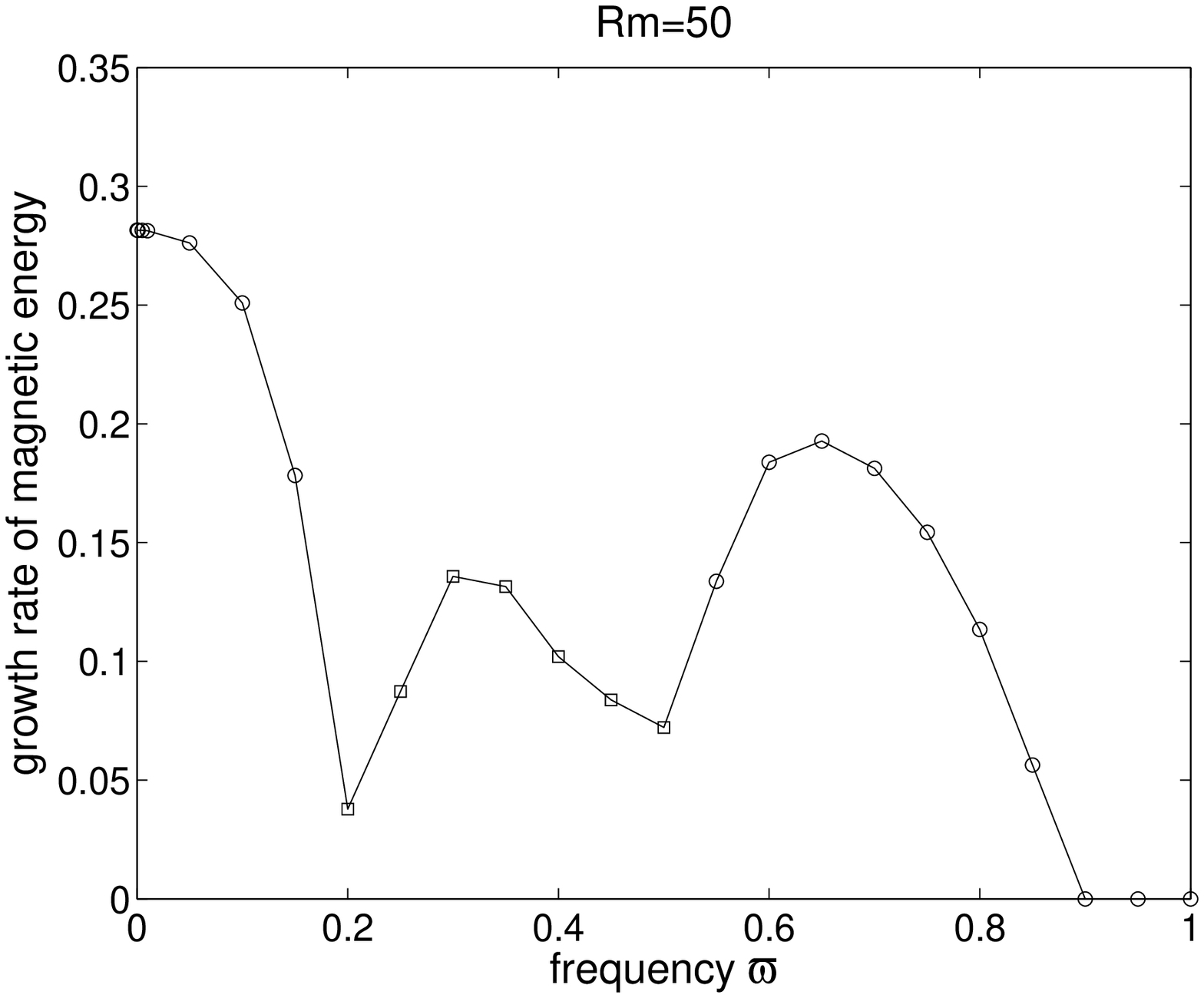}
\label{fig3b}}
\caption{Growth rate of magnetic energy against frequency $\varpi=\varpi_1=\varpi_2$ at $k_1=k_2=1$. (a) $Rm=20$ and (b) $Rm=50$. Circle symbol denotes $\sigma_i=0$ whereas square symbol denotes $\sigma_i\neq0$.}
\label{fig3}
\end{figure}

Next we test the effect of wave frequency on the growth rate. Figure \ref{fig3} shows the growth rate of magnetic energy against frequency. As the investigation of wavenumber, the two subfigures are at two different $Rm$'s. Generally speaking, a higher frequency corresponds to a lower growth rate, which is again qualitatively in agreement with (\ref{emf}). The dynamo fails when the wave frequency is high enough. The maximum frequency to sustain the dynamo action is higher at higher $Rm$. However, it is interesting that there exist some particular frequencies (square symbol) leading to the oscillations of magnetic energy, i.e., $\sigma_i\neq0$. The average of these oscillatory growth rates is plotted to result in the peaks on the curves. The structure of magnetic field is determined by $Rm$ and the wavenumber of helical wave but not the frequency of helical wave. Whatever frequency is, the magnetic energy is always on the large scale.

We then test some other cases. Firstly, we test dynamo induced by two helical waves with different frequencies. It shows that the growth rate of two different frequencies is between the two growth rates of two identical frequencies, e.g., $\sigma_r$ of $\varpi_1=0.01$ and $\varpi_2=0.05$ is between that of $\varpi_1=\varpi_2=0.01$ and that of $\varpi_1=\varpi_2=0.05$ (the other parameters are kept the same). Secondly, we test dynamo induced by two helical waves with different wavenumbers. Similar to the case of two different frequencies, the growth rate of two different wavenumbers is between those of two identical wavenumbers. The helicity of two helical waves with different wavenumbers is
\begin{align}
h&=(\bm u_1+\bm u_2){\,\bm \cdot\,}\bm\nabla\times(\bm u_1+\bm u_2)=(\bm u_1+\bm u_2){\,\bm \cdot\,}(k_1\bm u_1+k_2\bm u_2) \nonumber\\
&=(k_1+k_2)\bigl(u_0^2+\bm u_1{\,\bm \cdot\,}\bm u_2\bigr).
\end{align}
If $k_1+k_2=0$ then helicity vanishes. So thirdly, we test dynamo induced by two helical waves with opposite wavenumbers. It shows that the dynamo with zero helicity is quite difficult to be induced, i.e., the critical $Rm$ of the zero-helicity dynamo is much higher than that of the nonzero-helicity dynamo (with $k_1=1$ and $k_2=-1$ dynamo fails until $Rm=100$, and in contrast, with $k_1=k_2=1$ dynamo has occurred at $Rm=10$). This indicates that helicity is, though not necessary, helpful for the dynamo action (see equation (\ref{emf})).

To end this section we compare the $\alpha$-effect in the numerical calculation to that predicted by the first-order smoothing theory in both cases of an imposed field and of a self-excited dynamo. We firstly calculate the $\alpha$-effect induced by the interaction of one helical wave and an externally imposed uniform field. We use equation (7.62) in \cite{moffatt_book} for velocity, namely $\bm u=(\sin(kx_3-\varpi t),\cos(kx_3-\varpi t),0)$ and the uniform field is $\bm B_0=(0,0,1)$. The numerical result exactly conforms to the first-order smoothing theory, i.e., the e.m.f. $\bm u\times\bm b$ is a constant predicted by equation (7.70) in \cite{moffatt_book}. We next calculate the $\alpha$-effect induced by the interaction of three helical waves and an externally imposed uniform field. We use equation (9.47) in \cite{moffatt_book} for velocity, in which three stationary (zero-frequency) helical waves are superposed, namely $\bm u_1=(\sin kx_3,\cos kx_3,0)$, $\bm u_2=(0,\sin kx_1,\cos kx_1)$ and $\bm u_3=(\cos kx_2,0,\sin kx_2)$, and the uniform field is $\bm B_0=(1,1,1)$. The numerical result shows that the e.m.f. is not a constant but depends on position. This is because the e.m.f. consists of not only the auto-correlation terms such as $\bm u_1\times\bm b_1$, which induces a constant $\alpha$, but also the cross terms such as $\bm u_1\times\bm b_2$, which induces the $\alpha$ as a function of position. We finally calculate the $\alpha$ in the dynamo induced by the above three stationary helical waves. Because the first-order smoothing theory works better for smaller $Rm_{\rm wave}$, we choose $Rm=10$ and $k=40$ such that $Rm_{\rm wave}=0.25$, which is pushed to the limit of computational facility (lower $Rm_{\rm wave}$ requires higher $k$ and thus higher resolution). The theory predicts that $\alpha$ should be isotropic and its dimensionless value is $-Rm/k=-0.25$ as predicted by equation (9.49) in \cite{moffatt_book}. The numerical result shows that the spatially averaged magnetic field in the dynamo is zero and magnetic energy is mainly on the $K=0$ and $1$ modes. So we define the large-scale field $\bm B_0$ to consist of only the $K=0$ and $1$ modes. It should be noted that $\bm B_0$ is not uniform but a function of position. The e.m.f. is calculated as $\bm u\times\bm b=\bm u\times(\bm B-\bm B_0)$ and it is also a function of position. We then check whether all the three components at all the positions conform to the expression $(\bm u\times\bm b)_i/B_{0_i}=-0.25$ ($i=1,2,3$). This does not hold. The reason for this discrepancy could be that the first-order smoothing theory works in the regime of $Rm_{\rm wave}\ll 1$ but $Rm_{\rm wave}=0.25$ is far away from the regime where the first-order smoothing theory works. The $\alpha$-effect in the first-order smoothing theory is derived with the approximation that the scale of magnetic field is separable and the large scale field $\bm B_0$ can be regarded as uniform compared to small-scale flow $\bm u$ and field $\bm b$. However, if $Rm_{\rm wave}$ is not sufficiently small, then neither $\bm u$ nor $\bm b$ can be considered as having a sufficiently small scale, and moreover, $\bm B_0$ is no longer at a sufficiently large scale such that it cannot be considered as uniform.

\section{Discussion}
In this short paper we numerically investigate the kinematic $\alpha^2$-dynamo induced by two helical waves. This helical-wave dynamo is a slow dynamo. There exists an optimal wavenumber for the dynamo growth rate. Slower helical waves are better to the dynamo action. The oscillations of magnetic energy can be triggered at some particular wave frequencies. These results might give some hints for understanding the geophysical and astrophysical magnetic fields. The thermal Rossby waves in convection driven dynamo, the inertial waves in precession or tide driven dynamo, and the magneto-inertial waves in magneto-rotational instability driven dynamo, are all helical waves because of the presence of rotation. In these helical-wave dynamos, it can be inferred that the dynamo efficiency is the highest at some particular length and time scales of waves. It can be also inferred that in the Earth's core the slow magnetostropic waves are more powerful for the geodynamo than the fast inertial waves. This work is an numerical experiment which is expected to give some hint for the theoreticians to extend the theory of helical-wave dynamo, e.g., the asymptotic power law for the growth rate at $Rm\rightarrow\infty$. The further work is to involve the Navier-Stokes equation in the helical-wave dynamo to test the analytical results in \cite{moffatt_dynamic}.

\section*{Acknowledgement}
The two anonymous referees pointed out some mistakes in the first version and one of them suggested to compare the numerical result to the first-order smoothing theory. Prof.~Andreas Tilgner discussed with me about the optimal wavenumber. Dr.~C.K.~Chan helped me how to use FFT library. This work was supported by the National Science Foundation’s Center for Magnetic Self-Organization under grant PHY-0821899.

\bibliographystyle{gGAF}
\bibliography{paper}
\markboth{X.~Wei}{Kinematic dynamo induced by helical waves}

\end{document}